\documentclass[a4,center,fleqn]{article}

\usepackage{graphicx}
\usepackage{url}
\usepackage{xcolor}

\begin{document}

\title{Reweighting of molecular simulations with explicit-solvent SAXS restraints elucidates ion-dependent RNA ensembles}

\author{Mattia Bernetti$^1$, Kathleen B. Hall$^2$, and Giovanni Bussi$^1$}

\date{%
    $^1$Scuola Internazionale Superiore di Studi Avanzati, Via Bonomea 265, Trieste 34136, Italy\\%
    $^2$Department of Biochemistry and Molecular Biophysics, Washington University School of Medicine, St. Louis, Missouri 63110, USA\\[2ex]%
}

\maketitle

\begin{abstract}

Small-angle X-ray scattering (SAXS) experiments are increasingly used to probe RNA structure.
A number of \emph{forward models} that relate measured SAXS intensities and structural features,
and that are suitable to model either explicit-solvent effects or solute dynamics, have been proposed in the past years.
Here we introduce an approach that integrates atomistic molecular dynamics simulations and SAXS experiments to reconstruct RNA structural ensembles while simultaneously accounting for both RNA conformational dynamics and explicit-solvent effects. Our protocol exploits SAXS pure-solute forward models and enhanced sampling methods to sample an heterogenous ensemble of structures, with no information towards the experiments provided on-the-fly. 
The generated structural ensemble is then reweighted through the maximum entropy principle so as to match reference SAXS experimental data at multiple ionic conditions. Importantly, accurate explicit-solvent forward models are used at this reweighting stage. We apply this framework to the GTPase-associated center, a relevant RNA molecule involved in protein translation, in order to elucidate its ion-dependent conformational ensembles. We show that (a) both solvent and dynamics are crucial to reproduce experimental SAXS data and
(b) the resulting dynamical ensembles contain an ion-dependent fraction of extended structures.
\end{abstract}

\section{Introduction}

RNA molecules accomplish a plethora of functional roles in the cell, their function being dictated not only by their sequence and structure but also, to a large extent, by their dynamical behavior \cite{al2008rna,mustoe2014hierarchy}.
Molecular dynamics (MD) simulations have nowadays reached the status of a standard tool to explore the dynamics of biomolecular structure at the atomistic level \cite{dror2012biomolecular,sponer2018rna}.
Nevertheless, inaccuracies in the force fields and limitations in terms of the accessible timescales often make their capability to reproduce and to predict experimental results limited \cite{sponer2018rna}. The combination of MD simulations with experimental data is thus emerging as a robust asset to characterize the conformational dynamics of relevant biomolecules \cite{pitera2012use,bonomi2016metainference,bonomi2017principles,bottaro2018biophysical,cesari2018using}
including RNAs \cite{borkar2016structure,borkar2017simultaneous,kooshapur2018structural,paissoni2019martini,bottaro2020integrating,reisser2020conformational}.
Here, MD simulations can be seen as a powerful tool that complements experimental data making it possible to add dynamical information to experiments that report ensemble averages. This can be even more important when using low-resolution experiments such as small-angle X-ray scattering (SAXS) \cite{chen2016saxs,hub2018interpreting}.
Here, the synergy between MD and experiment allows faithful structural ensembles at atomistic resolutions to be generated \cite{kofinger2013atomic,chen2015structural,kooshapur2018structural,hermann2019saxs,paissoni2019martini,paissoni2020determination,jussupow2020dynamics,ivanovic2020small}.
SAXS experiments are particularly valuable in capturing the structural impact of changes in the ionic conditions,
that are highly relevant for RNA but poorly described by force fields \cite{sponer2018rna}.
Importantly, in the small-angle regime, most of the contribution to the overall SAXS is expected to originate from the solute.
Thus, from a computational standpoint, it is common practice to compute SAXS spectra using \emph{forward models}, \emph{i.e.}, equations to back-calculate the experiment from the simulated structures, based on the solute atomic coordinates only and including corrections to implicitly account for the solvent \cite{svergun1995crysol,schneidman2010foxs,nguyen2014accurate}. This choice reflects a compromise with the intensive effort that is typically involved to include the solvation explicitly in the computation. In recent years, methods have been devised that allow computing SAXS spectra including the solvent contribution through relatively efficient implementations, aiming at predicting SAXS spectra as accurately as possible
\cite{grishaev2010improved,kofinger2013atomic,knight2015waxsis,marchi2016first,hermann2019saxs,park2009simulated}.
This can be particularly critical when dealing with highly charged biomolecules, such as RNA, whose effect on the surrounding solvent and on the ionic cloud can be sizable up to a distance of several nanometers \cite{chen2009molecular,ivanovic2018quantifying}.
A possible route to combine MD and experimental data is to enforce the reference experimental data during the MD simulations. However, in the context of SAXS data, this option is hindered by practical limitations in the forward models. Indeed, whereas the on-the-fly estimate of the SAXS spectra is affordable for the pure solute, explicit solvent estimators have been used sparsely in this context \cite{hermann2019saxs}.

\begin{figure*}
\begin{center}
\includegraphics[width=\textwidth]{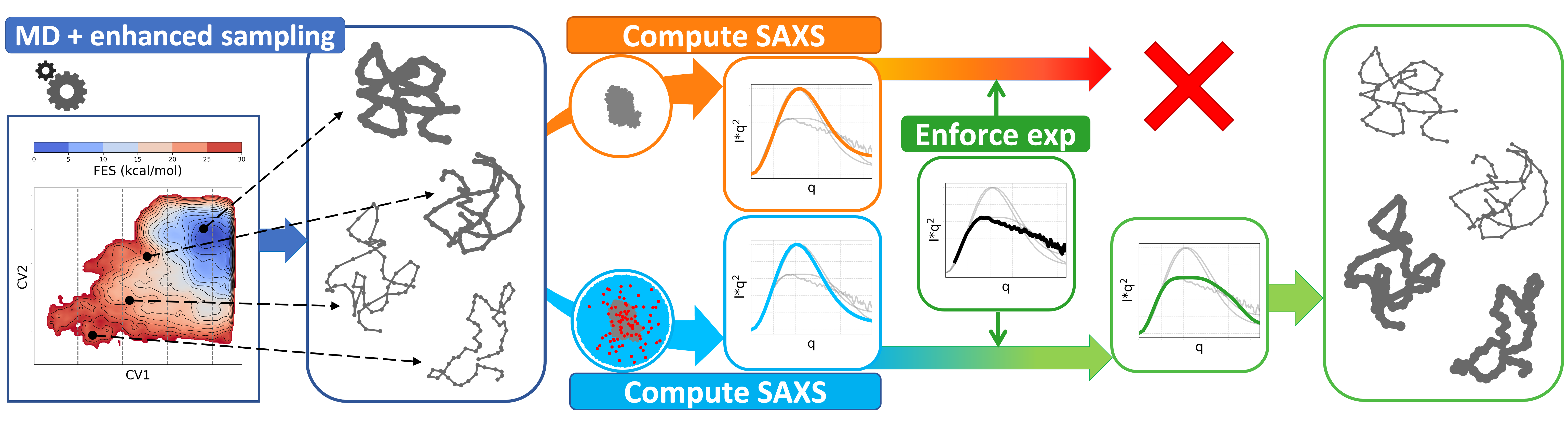}
\end{center}
\caption{Schematic pipeline of the protocol introduced in this work. Enhanced-sampling MD simulations with explicit solvent are initially employed 
	to sample RNA conformations of different structural compactness. At this stage, pure-solute forward models for SAXS are used as collective variables in a metadynamics protocol to guide the sampling, 
	with no information from experiments. We then compute \emph{a posteriori} the full SAXS spectra for the sampled structures, to be used as input in 
	the subsequent reweighting procedure. Notably, the SAXS spectra can either be computed from the pure solute only (top orange arrow) or including 
	the explicit solvent (bottom sky-blue arrow). 
	Experimental data are finally enforced through the maximum entropy principle to reweight the original ensemble (prior, blue rounded rectangle) and thus identify the least-modified ensemble 
	(posterior, green rounded rectangle) that is in agreement with the experimental reference. 
	Remarkably, we observe that the inclusion of the solvent in the computation of the SAXS spectra is a critical factor to achieve a successful 
	reweighting (bottom green arrow).
	}
\label{figure_1}
\end{figure*}

In this work, we introduce a protocol exploiting pure-solute forward models \cite{paissoni2019martini} and enhanced sampling \cite{mlynsky2018exploring} during MD simulations, to favor sampling of an heterogenous ensemble, without explicitly using the experimental data on-the-fly. The ensemble is then reweighted \cite{cesari2018using} to match experimental data using accurate explicit-solvent forward models \cite{kofinger2013atomic}.
We apply this framework to the GTPase-associated center (GAC), a 57-nucleotides-long RNA molecule of the 23S ribosomal subunit that is involved in protein translation \cite{moazed1988interaction,gao2009structure,weixlbaumer2008insights}. Recently, SAXS experiments reported on GAC structural flexibility in response to different ionic conditions in the buffer solution, noticing that Mg$^{2+}$ can stabilize the folded state, while K$^+$ favored less compact and more extended conformations \cite{welty2018divalent}. Through our protocol, we notice that explicit-solvent SAXS spectra are necessary to correctly reconstruct the ion-dependent structural ensembles and to obtain radii of gyration through Guinier fit that are compatible with the experiments (Figure \ref{figure_1}). In particular, in the case of K$^+$, we observed that a mixture of compact and extended structures is necessary to generate a structural ensemble that is in agreement with the experimental SAXS spectra.

\section{MATERIALS AND METHODS}

\subsection{Simulation details}
\begin{figure}
\begin{center}
\includegraphics[width=\columnwidth]{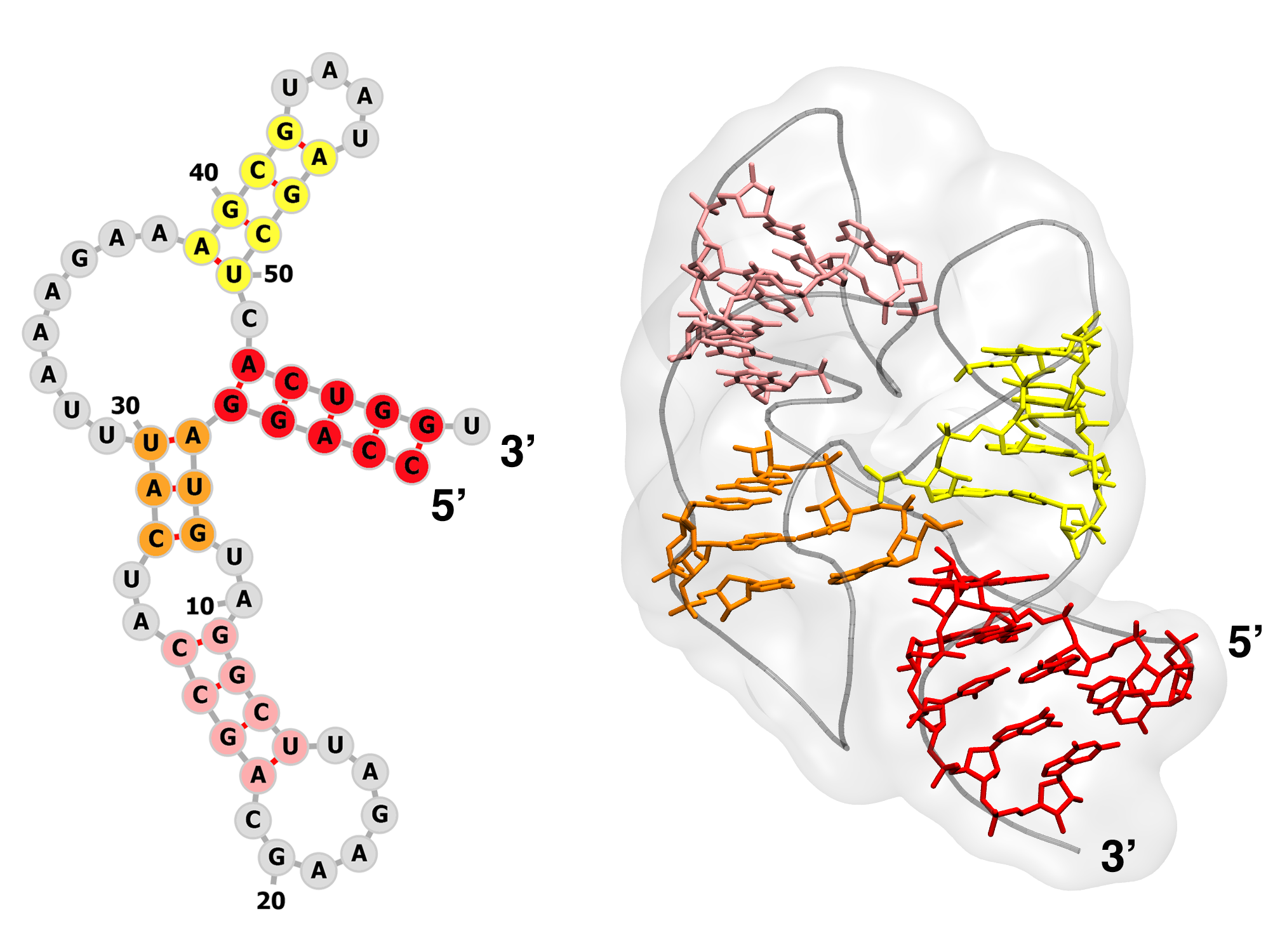}
\end{center}
\caption{GAC RNA structure. The stem regions are consistently color coded in the secondary (left panel) and tertiary (right panel) structure 
	representation. In the tertiary structure (PDB ID: 1HC8), the RNA backbone is represented as tube, stem heavy atoms are shown in licorice and the 
	overall shape of the molecule in depicted in shaded gray.
	}
\label{figure_2}
\end{figure}

All systems were prepared from the available crystal structure of GAC in its folded state (PDB ID: 1HC8 \cite{conn2002compact}, see Figure \ref{figure_2})
after removing the bound protein. We notice that a recent RNA-only crystal structure is virtually identical \cite{welty2020ribosomal}.
The system was described using the AMBER force field for nucleic acids \cite{cornell1995second,perez2007refinement,zgarbova2011refinement}, the 4-point optimal-point-charge (OPC) model for water \cite{izadi2014building}, and compatible ion parameters \cite{Joung2008determination,allner2012magnesium}.
4-point water models have been reported to improve the accuracy of simulated hydration effects in molecular systems.
This is particularly critical for biomolecules exploring heterogeneous ensembles including more extended structures, where an accurate representation of solute-water interactions is crucial to avoid overly compact conformations
\cite{best2014balanced,palazzesi2015accuracy,shabane2019general,hermann2019saxs}.
The OPC model has been shown to improve the agreement of simulation and experiment for unstructured RNA tetramers
\cite{bergonzo2015improved,bottaro2018conformational}.
All simulations were performed using GROMACS 2018.4 \cite{abraham2015gromacs}.
Four ionic conditions were considered with an increasing number of Mg$^{2+}$ ions, namely: 1. none, 2. crystallographic Mg$^{2+}$, 3.
crystallographic Mg$^{2+}$ plus half the amount needed to neutralize the system and 4. crystallographic Mg$^{2+}$ plus the amount needed to
neutralize the system (a representative sketch of the 4 systems is provide in Figure \ref{figure_3}a). In all cases, the crystal K$^+$ was retained, a 100 mM concentration of KCl was set for the buffer, and neutralization, where needed, was achieved by adding K$^+$ accordingly. 
The systems were initially minimized applying soft position restraints on RNA and water heavy atoms and on the crystal ions.
A multi-step equilibration was then conducted: 3 short simulations lasting 200 ps at 100, 200 and 300 K in the NVT ensemble using the velocity-rescaling thermostat \cite{bussi2007canonical}, with soft position restraints on RNA heavy atoms and on crystal ions; in further 900 ps in NPT ensemble using the Parrinello-Rahman barostat \cite{parrinello1981polymorphic}, these restraints were gradually removed to relax the system. Production runs were performed in the same ensemble.
Short unbiased MD simulations (10 ns) were run using a large rhombic dodecahedron simulation box containing approximately 170000 atoms, with edges distant 3 nm from all RNA atoms, using soft position restraints on solute heavy atoms to exclude solute dynamics effects. Longer unbiased MD simulations (1 $\mu$s) were performed using a smaller rhombic dodecahedron simulation box containing approximately 100000 atoms,
with edges distant 2 nm from all RNA atoms, with no position restraints, for the ionic conditions 1 and 4 (Figure \ref{figure_3}a) described above.

\subsection{Enhanced sampling}
Enhanced sampling MD simulations \cite{mlynsky2018exploring} were conducted on the system with no Mg$^{2+}$ (Figure \ref{figure_3}a) using the larger rhombic dodecahedron simulation box (approximately 170000 atoms). The Hamiltonian replica exchange (HREX) method~\cite{sugita2000replica} with scaling applied to selected residues \cite{bussi2014hamiltonian} (non-stem residues, Figure \ref{figure_2}, left panel in grey) was used to relax the contacts of regions outside of the stems. The lambda scaling factors were in the range 0.7--1.0 and distributed on 16 replicas according to a geometric progression. To ensure charge neutrality at each replica, the missing charge was spread over the dummy atom of all water molecules. Exchanges between neighboring replicas were attempted every 400 MD steps. HREX was combined with metadynamics \cite{laio2002escaping,camilloni2008exploring,bussi2020using}, where two
collective variables (CVs) were biased: the Ratio between peak ($q=0.1$ \AA$^{-1}$) and shoulder ($q=0.2$ \AA$^{-1}$) of the pure-solute SAXS spectrum in the Kratky form computed using the MARTINI model \cite{paissoni2019martini} and an additional variable (Diff) purposely designed to estimate the degree of formation of tertiary contacts in the RNA molecule.
Diff is defined as the root mean square of the G-vectors introduced in \cite{bottaro2014role} and can be computed
in practice taking the root-square difference of the eRMSD \cite{bottaro2014role,bottaro2016free} with respect to an arbitrary structure with no contacts formed,
computed using either the whole sequence or only the stem regions (see Figure S1).
Gaussians with height 2.09 kJ mol$^{-1}$ and $\sigma$ of 0.035 and 0.05 were deposited every 400 steps with a bias factor of 10 \cite{barducci2008well}.
In order to decrease the computational overhead of computing collective variables, the bias was applied every second step \cite{ferrarotti2015accurate}.
Stem regions were restrained setting an upper wall on the eRMSD with respect to the initial state at 0.7 using a force constant of 41.84 kJ mol$^{-1}$.  Upper (at 25 \AA) and lower (at 16 \AA) wall restraints were also applied to the $R_g$ with force constant 41.84 kJ mol$^{-1}$ nm$^{-2}$. Finally, an upper wall on the Ratio CV was placed at a value of 2.7 with a force constant of 41.84 kJ mol$^{-1}$. The whole setup was achieved through the open-source, community-developed PLUMED library \cite{tribello2014plumed,bonomi2019promoting} version 2.5. The simulation was conducted in the NVT ensemble for 180 ns/replica. Weights were computed a posteriori using the final bias \cite{branduardi2012metadynamics}. Representative sampled structures were selected taking advantage of the Quality Threshold (QT) clustering algorithm \cite{gonzalez2019quality}.

\subsection{Backcalculation of SAXS spectra}
The SAXS spectra for all the structures sampled during the MD simulation were computed in the $q$ range 0--0.5 \AA$^{-1}$ with a pace of 0.01 \AA$^{-1}$. Pure solute spectra for the short (10 ns), long (1 $\mu$s), and enhanced sampling simulations were computed on the all-atom structures with PLUMED, relying on a MARTINI bead representation of the system as recently introduced \cite{paissoni2019martini}. Notably, the method has been shown to produce coincident results in the small angle regime
that is relevant here (0$<q<$0.3 \AA$^{-1}$)
when compared with the all-(solute)-atom calculation (see also Figure S2).
Crysol spectra were computed for the short simulations using a maximum order of harmonics of 20 and default parameters, with the software version 2.8.4 \cite{svergun1995crysol}. Waxsis spectra were computed for the short simulations with an envelope constructed at a distance of 7 \AA{} from the RNA molecule and using default parameters, as implemented in the modified version of GROMACS 4.6.2 \cite{knight2015waxsis}. To compute the SAXS spectra with Waxsis, 1000 frames from the 10 ns simulations were used, along with 1000 frames from an independent pure-solvent simulation with the same salt concentrations and lasting 5 ns. Capriqorn spectra were computed for the short (10 ns), long (1 $\mu$s), and enhanced sampling simulations through the method introduced by Köfinger and Hummer \cite{kofinger2013atomic} implemented in the Capriqorn software (https://github.com/bio-phys/capriqorn) version 1.0.0. The method first computes separately the radial distribution function from a pure-solvent simulation with the same salt concentrations, for which purpose we used the same 5 ns pure-solvent simulations as for Waxsis. For the short and long simulations, a sphere geometry with radius 40 \AA{} was used along with a shell for solvent matching of width 7 \AA; for the enhanced sampling simulations, a sphere radius of 54 \AA{} and shell width of 10 \AA{} was employed (see Figure S3 for a schematic depiction).
$R_g^*$ was computed from the SAXS spectra through the standard Guinier fit procedure. Specifically, the linear fit was conducted in the $q$ range 0.02--0.06 \AA$^{-1}$ of the SAXS spectra in the $\ln I$ vs $q^2$ form, using scipy \cite{virtanen2020scipy}; $R_g^*$ is then computed as $\sqrt{-3m}$, where $m$ is the slope of the fitted line.

\subsection{Ensemble reweighting}
We then used ensemble reweighting to enforce the experimental spectrum for four chosen values of $q$.
Two values ($q=$0.03 and $q=$0.05 \AA$^{-1}$) were chosen inside the range used to compute $R_g^*$ through Guinier fit, while other two values ($q=$0.1 and $q=$0.2 \AA$^{-1}$) correspond to peak and shoulder of the SAXS spectrum in the Kratky form, respectively, which are more sensitive to variations in the structural compactness of the RNA solute.
The standard maximum-entropy reweighting procedure \cite{pitera2012use,cesari2018using} consists in determining a set of Lagrangian multipliers
$\lambda_i$ so that averages computed along the entire trajectory with weights proportional to $w_t=e^{-\sum_i \lambda_i s_i(t)+\frac{V_b(q(t))}{k_BT}}$
are equal to the reference values $s_i^{exp}$.  Here $s_i(t)$ is the value of the $i$-th observable in the $t$-th frame,
$V_b(q(t))$ is the metadynamics bias at the final time recomputed for the coordinates of the $t$-th frame, and $k_BT$ the thermal energy.
However, when used to directly enforce SAXS intensities, this procedure requires
an arbitrary scaling factor to be fixed \emph{a priori} or estimated during the analysis \cite{paissoni2020determination}.
To overcome this issue, we here enforce the weighted average of the quantity
\begin{equation}
s(q,t)=I(q,t)-\frac{I^{exp}(q)\sum_{q'} I(q')}{\sum_{q'} I^{exp}(q')}
\end{equation}
to be equal to zero,
where $I(q,t)$ is the SAXS intensity at wavevector $q$ for the $t$-th frame.
By straightforward manipulation it can be seen that $\langle s(q,t) \rangle=0$ implies
$\langle I(q,t) \rangle \propto I^{exp}(q)$.
Since by construction $\sum_i s(q,t)=0$, the resulting ensemble is invariant with respect to a uniform shift of the
four multipliers $\lambda_i$. It is then possible to arbitrarily set their sum to zero,
\emph{i.e.} $\sum_i\lambda_i=0$.
The reweighting was performed using a custom python script taking advantage of scipy minimization algorithms \cite{virtanen2020scipy}.
Two sets of experimental datapoints reported in Ref.~\cite{welty2018divalent} were used to reweight our samples:
one set was obtained using 100 KCl concentration and another using an additional 1mM MgKCl concentration. Both experiments also included 10mM MOPSO as a buffer.

Since weights can be uneven, only the structures with the highest weights effectively contribute to the ensemble.
An approximate estimate of the number of structures with a significant weight can be obtained by computing the
Kish effective sample size, defined as $K=\left(\sum_t w_t\right)^2/\left(\sum_t w_t^2\right)$ \cite{gray1969survey}.

\section{RESULTS}

\subsection{Spectra from MD simulations of crystal structure}
\begin{figure}
\begin{center}
\includegraphics[width=0.8\columnwidth]{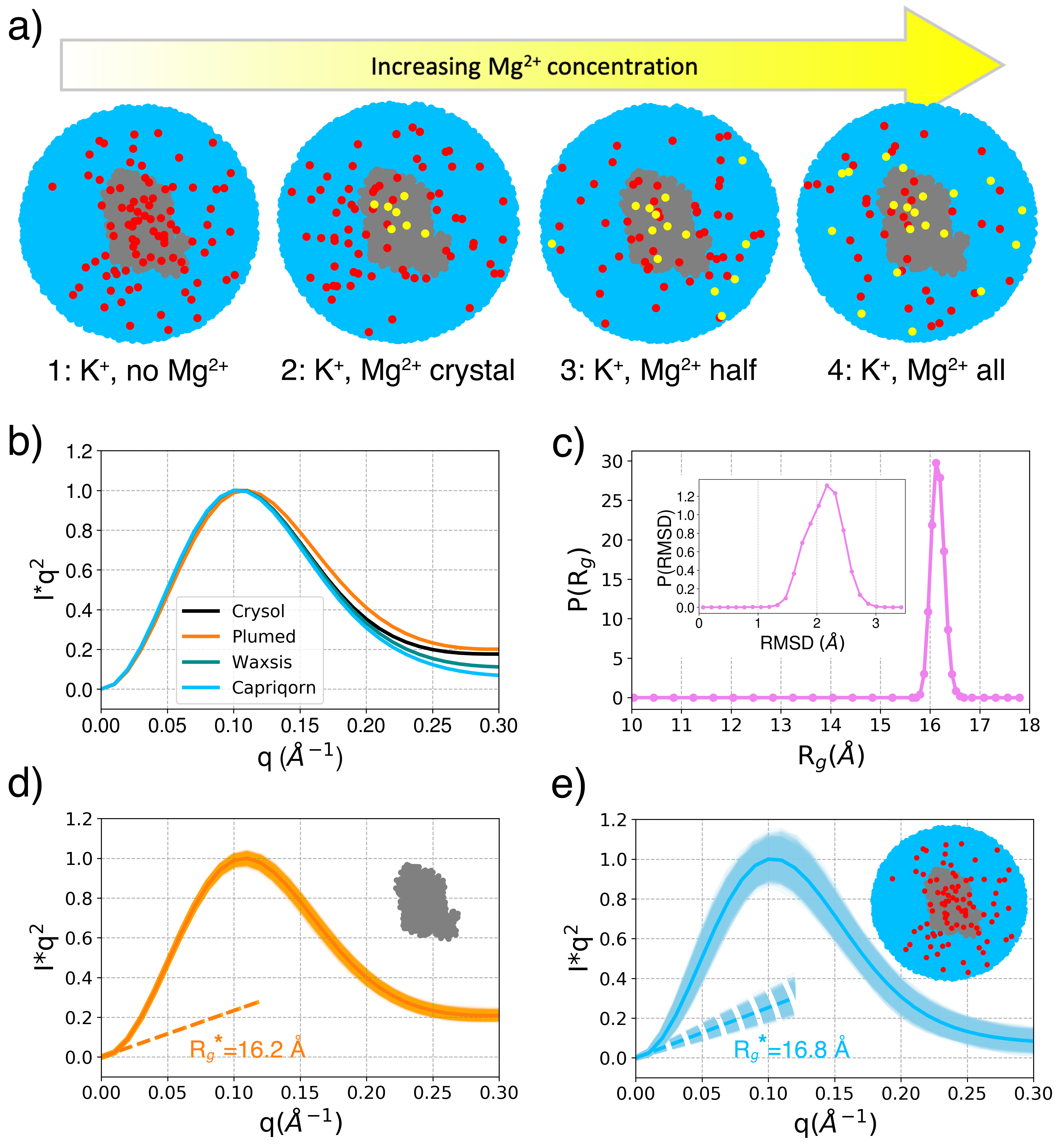}
\end{center}
	\caption{Simulated buffers and SAXS spectra from unbiased MD. a) Schematics representing the four tested setups.
                 All setups were used in short (10 ns) simulations with restrained solute. Setups 1 and 4 were also used in longer (1$\mu$s) 
                 unrestrained simulations. Setup 1 was used in enhanced sampling simulations.
  b) Comparison of SAXS spectra computed using different available methods on a 10 ns simulation (K$^+$ only, no Mg$^{2+}$)
	with soft restraints on RNA 
	heavy atom positions. The spectra are displayed in the Kratky form as $Iq^2$ vs $q$ and normalized by their peak ($q=0.1$ \AA$^{-1}$) values. 
	c) Probability density function for the geometric gyration radius $R_g$ and 
	RMSD from native for a 1 $\mu$s long unbiased MD simulation, with 
	explicit solvent, started from the crystal structure (PDB ID: 1HC8). d) SAXS spectrum for the 1 $\mu$s unbiased MD run computed from the pure solute. 
	The frame-by-frame spectra are shown as a light orange shade and display a root-mean-square deviation from the resulting average, in orange, of 0.01 in the displayed units. The colored dashed line indicates 
	the Guinier fit performed at low-$q$ values of the average spectrum to compute the effective gyration radius$R_g^*$. In the top right corner of the plot, the small sketch indicates 
	that the spectra are computed from the pure solute (RNA, in gray) only. e) SAXS spectrum from the 1 $\mu$s unbiased MD run computed including the explicit solvent. 
	The frame-by-frame spectra are shown as light sky-blue shade and display a root-mean-square deviation from the resulting average, in sky-blue, of 0.03 in the displayed units. In the top right corner, the small 
	sketch indicates that the spectra are computed from the whole system (RNA, in gray, water, in sky-blue, and ions, in red).
Spectra in panels d) and e) are normalized by the peak of the average spectrum.
	}
\label{figure_3}
\end{figure}

Short unbiased MD simulations, lasting 10 ns, were performed with varying ionic conditions (see Figure \ref{figure_3}a) on GAC RNA starting from the crystal structure, where GAC is in its folded state. Specifically, in the four considered setups, the concentration of Mg$^{2+}$ ions was gradually increased. Soft position restraints were placed on the RNA heavy atoms to exclude relevant solute dynamics while guaranteeing adequate sampling for the solvent. SAXS spectra were then computed from the MD sampled structure through different available software, namely PLUMED \cite{tribello2014plumed,paissoni2019martini}, Crysol \cite{svergun1995crysol}, WAXSiS \cite{knight2015waxsis} and Capriqorn \cite{kofinger2013atomic}. A substantial difference between these methods is how they include the contribution of solvent to the overall SAXS spectra (see Figure S4 for a schematic depiction). In particular, while PLUMED purely relies on the solute coordinates and CRYSOL models the solvent implicitly, both WAXSiS and Capriqorn include explicitly the solvent species in the computation of the spectra. Consistently, a comparison of the predicted SAXS spectra (Figure  \ref{figure_3}b) for a same simulation displays differences between the methods, which are more marked in the region at $q>$ 0.1 \AA$^{-1}$ where the contribution of the solvent becomes more significant. Most notably, WAXSiS and Capriqorn, both including explicitly the solvent contribution, provide compatible results. Additionally, no remarkable dependence on the different ionic conditions explored was observed when comparing results from the different setups (see Figure S5).

To investigate the role of RNA dynamics and potentially observe extended conformations, we performed long unbiased MD simulations of 1 $\mu$s with no restraints (setups 1 and 4 in Figure \ref{figure_3}a). Interestingly, no remarkable structural changes were observed on this time scale (see Figure \ref{figure_3}c), although the simulation with a pure K$^+$ buffer had a slightly larger displacement from the crystal structure and a slightly larger
gyration radius than the simulation including Mg$^{2+}$  (see Figure S6).
Average SAXS spectra from the 1 $\mu$s simulations were then computed using the pure solute (Figure \ref{figure_3}d) or including the solvent contribution (Figure \ref{figure_3}e). It is worth noticing that, when using a method that explicitly takes into account the solvent contribution (Figure \ref{figure_3}e), the frame-by-frame spectra exhibit significant variations from the spectrum obtained by averaging over the whole trajectory. On the contrary, when a pure-solute method is employed (Figure \ref{figure_3}d), the frame-by-frame spectra are closer to their average. Consistently, the fluctuations in the corresponding $R_g^*$, as computed through the Guinier fit procedure (see Materials and Methods) in the low-$q$ values of the SAXS spectra, are wider when the solvent is included explicitly. Standard deviation of $R_g^*$ was 0.1 \AA{} and 1.3 \AA{} for the pure solute and explicit solvent calculations, respectively. Most notably, for the same simulation, a greater value of $R_g^*$ as computed from the full simulation is observed when including the solvent, namely 16.8 \AA{} from SAXS with solvent vs 16.2 \AA{} from pure-solute SAXS, with statistical errors $<0.1$ \AA{}.

\subsection{Spectra from enhanced-sampling MD simulations}

In order to reproduce SAXS experimental spectra associated with diverse ionic conditions, enhanced sampling methods \cite{mlynsky2018exploring} were used to generate conformations with a broad range of compactness. Representative sampled structures are given in the supplementary material. Figure \ref{figure_4} reports the free-energy surfaces (FES) reconstructed from enhanced sampling simulations started from the crystal conformation. The system was biased with metadynamics through one variable that favored disruption of tertiary contacts and another that guided the sampling towards more extended structures and relaxation was enhanced by a replica exchange protocol (see Materials and Methods). We stress here that no information to guide towards the experiment was used in the sampling procedure and that only the no-Mg$^{2+}$ setup (setup 1 in Figure \ref{figure_3}a) was simulated. The resulting FES displayed a unique global minimum in correspondence of the crystal conformation (star label in the leftmost panel of Figure \ref{figure_4}). Correspondingly, more extended conformations were located in higher free-energy regions, up to about 25 kcal mol$^{-1}$ from the minimum. SAXS spectra computed \emph{a posteriori} using sets of frames with different compactness are also reported, both from the pure solute and the explicit solvent calculation (Figure \ref{figure_4}, panels with SAXS spectra).
For FES regions where GAC structure was significantly expanded compared to the initial state of the simulation, the shape of the SAXS spectra in the Kratky form changed remarkably. In all cases, a significant difference can be observed between the SAXS spectra computed from the pure solute and including the solvent explicitly (Figure \ref{figure_4}, orange and sky-blue lines in the SAXS spectra). 

\begin{figure*}
\begin{center}
\includegraphics[width=\textwidth]{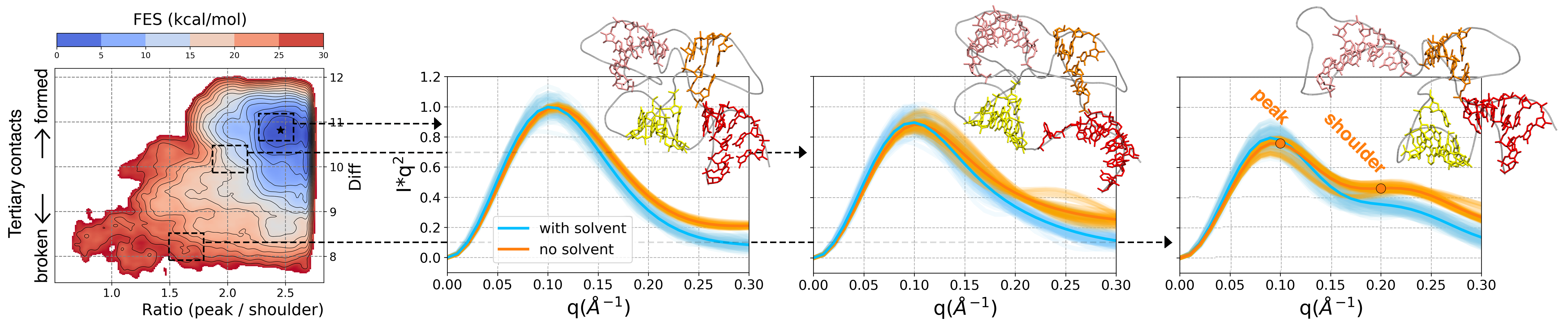}
\end{center}
\caption{Free-energy surface (FES) and SAXS spectra from the enhanced-sampling MD simulation. The FES is shown as a function of the biased collective variables, namely the ratio 
	between peak ($q=0.1$ \AA$^{-1}$) and shoulder ($q=0.2$ \AA$^{-1}$) of the pure-solute SAXS spectrum in the Kratky form, and a variable (Diff) estimating the degree of formation of 
	tertiary contacts in the RNA molecule (see the methods section for its definition). The star indicates the crystal structure, from 
	which the simulation was initiated. For representative regions of the FES (dashed rectangles), the corresponding SAXS spectra are shown (panels indicated by the dashed arrows). 
	Specifically, both the average spectrum computed from the pure solute (RNA only) and from the whole system (RNA and solvent) are displayed in orange and 
	sky-blue, respectively. The corresponding frame-by-frame spectra are also reported as light orange and light sky-blue shades. From left to right panels, their root-mean-square deviations from the corresponding averages are 0.01, 0.07, 0.14 for the pure solute and 0.03, 0.06, 0.12 for the whole system, in the displayed units. For each region, a representative structure of GAC is shown, color 
	coded consistently with Figure 2.
	}
\label{figure_4}
\end{figure*}

\subsection{Enforcing experimental spectra}
\begin{figure}
\begin{center}
\includegraphics[width=\columnwidth]{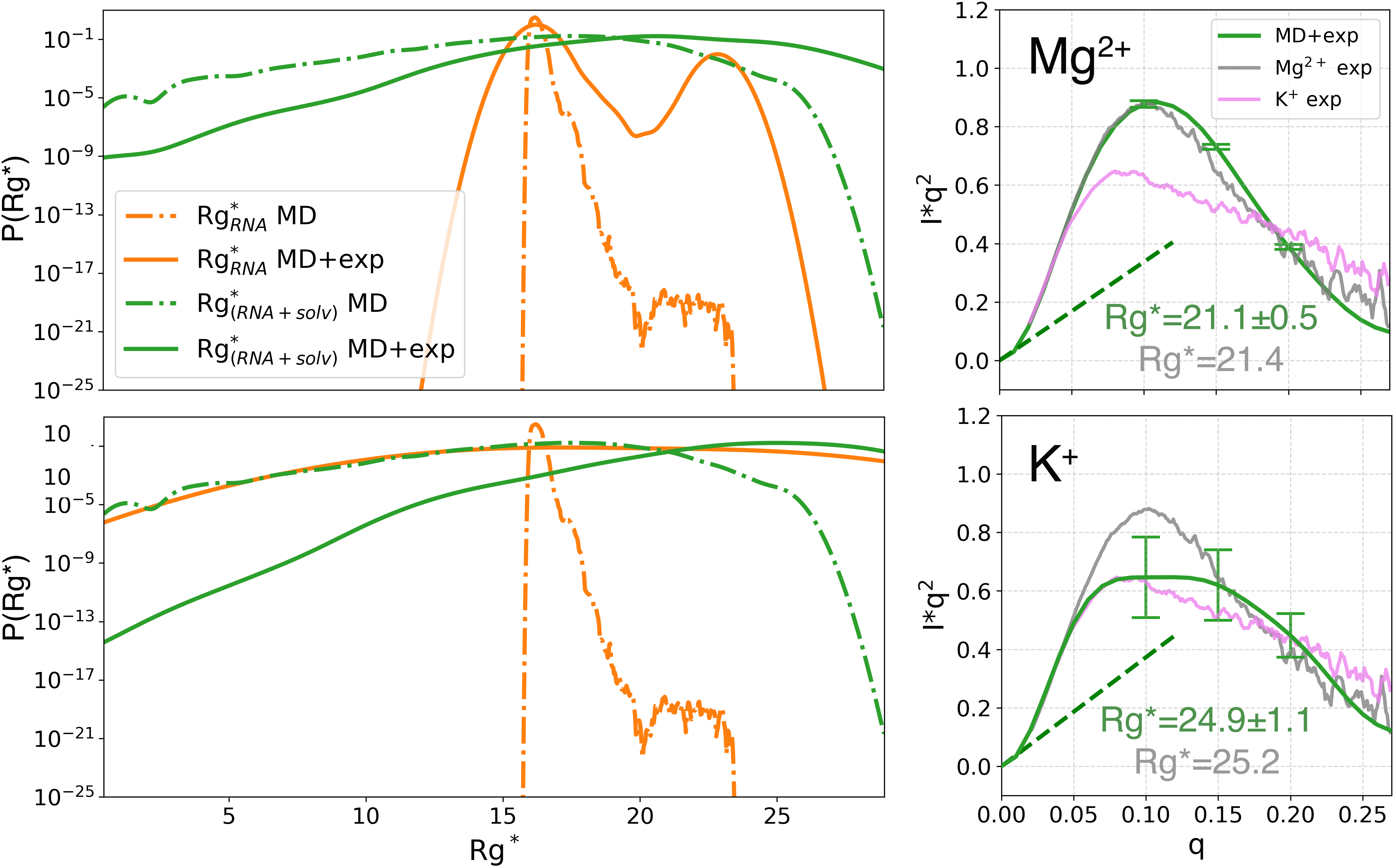}
\end{center}
	\caption{Probability distribution of $R_g^*$ and SAXS spectra from the
        ensembles reweighted to match Mg$^{2+}$ experimental data (top panels) and K$^+$ experimental data (bottom panels).
        In the left panels, $R_g^*$
	is computed through a Guinier fit on the SAXS spectra of the original (prior) ensemble (dashed lines) and of the reweighted (posterior) ensemble 
	(solid lines). Results for both the pure solute (orange lines) and the whole system (green lines) are shown. 
	Note that both solid lines are based on the same reweighted ensemble, \emph{i.e.} the one matching experiments and SAXS spectra computed including the solvent. In the 
	right panels, the SAXS spectra from the reweighted ensembles are shown in the Kratky form and also reported are the corresponding $R_g^*$ and the associated statistical error 
	(in green). The statistical errors for selected $q$ values of the spectra are displayed as error bars. The reference experimental value for $R_g^*$, which was 
	computed from the experimental SAXS spectra (in light gray and violet for Mg$^{2+}$ and K$^+$, respectively), is also reported in gray for comparison.
	}
\label{figure_5}
\end{figure}

We conducted a reweighting procedure to possibly identify GAC structural ensembles underlying reference experimental SAXS spectra among those collected in the enhanced sampling simulation. As a first step, the SAXS spectra were computed for each of the obtained structures. Importantly, the spectra were computed for both the pure solute and for the whole system, \emph{i.e.}, with the explicit inclusion of the solvent (Figure \ref{figure_4}, orange and sky-blue lines). Subsequently, a reweighting procedure was conducted using the maximum entropy principle \cite{pitera2012use,cesari2018using}. In particular, the ensemble generated in the K$^+$-only simulation was reweighted so as to match experimental SAXS spectra obtained in presence of Mg$^{2+}$ or only K$^+$  (Figure S7). To this end, the agreement with the experiment was enforced in four $q$ points of the spectrum that included the Guinier fit region and the peak and shoulder points of the Kratky plot. In this way, both the $R_g^*$ and the shape of the Kratky plot were taken into account. Interestingly, while employing as prior observables the spectra computed with the contribution from the solvent allowed to identify posterior ensembles consistent with the experimental ones (Figure S8), this was not the case when performing the procedure using the pure-solute spectra (Figure S9). This can be explained by the different distributions of the $R_g^*$ in the prior ensemble (Figure \ref{figure_5}, left panels). While $R_g^*$ values were confined in the range 15-24 \AA{} in the pure-solute case, the range explored was broader when the spectra included the solvent contribution.
In other words, despite they were applied to the same pool of structures, the two different ways of computing the spectra (pure-solute vs solute + solvent) produced different effects in terms of the resulting $R_g^*$. In particular, $R_g^*$ was greater than the corresponding $R_g$ computed from the only solute coordinates, with the effect being more marked on compact structures and becoming milder for more extended ones (Figure S10).

The refined distribution obtained when enforcing the Mg$^{2+}$ experimental data revealed a prevalence of compact structure complemented by a small population (about 1\%) of extended structures.
The population of extended structures was significantly larger (about 42\%) when the refinement was done using K$^+$ experimental data
(Figure \ref{figure_5}, upper and lower left panels, respectively).
It is important to notice that this reweighting stage resulted in very low Kish sizes \cite{gray1969survey,rangan2018determination} (15.9 and 2.9 for Mg$^{2+}$ and K$^+$, respectively, to be compared with 36000 structures used in the analysis), indicating that only a very limited number of structures contributed to the final spectrum. This is reflected in a high statistical error for the final spectrum, as it can be appreciated in Fig.~\ref{figure_5},
especially when using K$^+$ experimental data. This effect can be ascribed to a limited sampling of more extended GAC conformations in the enhanced sampling simulations.
The corresponding reweighted SAXS spectra were compatible with the experimental reference for Mg$^{2+}$ and K$^+$ (Figure \ref{figure_5}, upper and lower right panels, respectively). As expected based on the low Kish size, the agreement was poorer for the case of K$^+$. Nevertheless, the predicted $R_g^*$ values were in agreement with the experimental ones (compare green and gray values in the right panels of Figure \ref{figure_5}).

We also tested if adding a rigid shift to the experimental data could increase the Kish size or result in a Guinier radius in better agreement with experiment \cite{schneidman2013accurate}. The analysis suggests that no shift is necessary (see Figure S11).

\section{DISCUSSION}

In this work we use atomistic molecular dynamics simulations, coupled with enhanced sampling methods, to generate a conformational ensemble for GAC RNA. The resulting ensemble is then reweighted so as to enforce agreement with recently published SAXS data \cite{welty2018divalent}. SAXS spectra are back-calculated comparing approaches that include solvent effects to a different extent. Preprocessed data and a notebook that can be used to generate all the figures can be downloaded at \url{https://github.com/bussilab/saxs-md-gac}. Full trajectories are available at \url{https://doi.org/10.5281/zenodo.4646262}.

The introduced approach combines two crucial ingredients, namely conformational dynamics and solvent effects. The former is enhanced by using replica exchange and metadynamics acting on a proxy of the SAXS intensities, so as to encourage the exploration of extended structures. The latter is achieved by using an accurate \textit{a posteriori} calculation of the spectra, where the presence of the solvent is explicitly modeled \cite{kofinger2013atomic}, in conjunction with the maximum entropy principle. The approach is summarized in Fig.~\ref{figure_1}. We note here that all the steps involved in the procedure are conducted through freely available software.
When enforcing experimental data obtained in presence of Mg$^{2+}$, our procedure suggests a low population of extended structures to contribute to the spectrum. Albeit low, this contribution is necessary to explain the difference observed between the experimental SAXS data and the spectra back-calculated using the explicit solvent approach on the native structure. When enforcing experimental data obtained in absence of Mg$^{2+}$, instead, the fraction of extended structures is significantly higher.

We first addressed the role of solvent contribution to the SAXS intensities. We thus ran long MD simulations starting from the crystal structure and analyzed them using a range of methods that neglect solvent contribution or include it at different levels of approximation and compare the back calculated SAXS spectra with the available experimental data. From our results it emerges that (a) an explicit inclusion of the solvent has a measurable effect on the apparent radius of gyration of the molecule as obtained with the Guinier fit and (b) the overall shape of the spectrum is affected also at relatively small scattering vectors ($q<0.3$\AA$^{-1}$). Although none of the employed methods results in a SAXS spectrum in agreement with the available experimental data, irrespective of the ionic condition in which the experiments were performed and the simulations were run, the explicit-solvent methods lead to values of the observed gyration radius that were closer to the experimental ones. This suggests that the role of solvent contribution in the estimation of SAXS spectra for nucleic acids should not be neglected. We hypothesize this to be related to the high charges that make the solvation shell structured and of critical importance.

We then addressed the role of RNA dynamics. Unbiased MD simulations in the $\mu$s timescale stably maintained the compact, crystal conformation of GAC and no alteration of the structure towards more extended states was observed, irrespective of the simulated ion conditions. This suggested that the disruption of the initial conformation would require remarkably extensive sampling. Indeed, the crystal structure comprises a complex network of tertiary contacts between the four separate stem regions of GAC, which overall results in a significant structural stability. 
We thus employed a combination of two enhanced sampling methods, namely metadynamics and replica exchange. The former was used to improve the sampling of extended conformation by biasing dedicated CVs. Specifically, a first CV was designed to quantify the amount of tertiary contacts in the structure, so as to facilitate its rupture and to explore different sets of contacts, while a second CV was designed to reflect salient features of the SAXS spectra (peak and shoulder of the Kratky form), so as to encourage the exploration of structures with heterogeneous spectra and thus of varying structural compactness. At the same time, the replica exchange method was exploited to soften energetic barriers, thus accelerating local reconformations. 

It is important to underline that no experimental data were used while performing MD simulations, and that SAXS intensities used for the metadynamics simulations were estimated using a pure-solute coarse-grain approach, that can be effectively used on-the-fly. This was possible here since the estimated intensities were only used to enhance sampling, and not to reproduce the experimental data. We notice that also in Ref.~\cite{kimanius2015saxs} SAXS intensities calculated in a pure-solute approximation were used to enhance sampling. Interestingly, also in the extended structures we found discrepancies between the pure-solute calculation of the SAXS intensities and their explicit-solvent counterparts. The effect of the solvent is not trivial and cannot be reduced to an increased effective gyration radius. In fact, the estimated Guinier radius for extended structure appeared not to be dependent on the inclusion of the solvent.

Although enhanced sampling simulations were capable to generate extended structures with SAXS spectra compatible with experiment, the  weight of these structures in the Boltzmann ensemble was very low. This happened even if our simulations were performed in absence of Mg$^{2+}$, and thus in conditions that are expected to favor a significant population of extended structures. This can be explained to be a consequence of (a) the choice of initializing our simulations from a compact structure and (b) a possible bias in the employed force field towards compact structures. Only by using a reweighting procedure based on the maximum entropy principle we were able to reproduce the experimental spectra.
The standard maximum-entropy-based reweighting was here extended so as to allow fitting experimental data that are known up to an a priori undetermined pre-factor.
Notably, the employed maximum-entropy reweighting only succeeded when spectra were calculated including the solvent, confirming that both solvent contribution and dynamics are crucial in this system. We observe that it is possible to extend the maximum entropy principle so as to model experimental errors by including a regularization term \cite{cesari2018using}. Error models are also explicitly included in the metainference method \cite{bonomi2016metainference}. For simplicity, we didn’t include any regularization or error model here, since the statistical error on the intensities due to the limited sampling dominates on the experimental error. When attempting a fit based on the pure-solute spectra, instead, adding a regularization term was required for the minimization to converge but then lead to spectra in remarkable disagreement with experiment.

The decision to avoid an explicit usage of the available SAXS data during the simulation was here made so as to generate a generic ensemble that spans a wide range of conformations and can be then, \emph{a posteriori}, reweighted in order to match experimental data obtained with different ionic conditions. In this specific case, it allowed for obtaining reference ensembles for two different datasets from the same simulation, thus at half of the computational cost. Using the same simulation for fitting two separate experiments has also the advantage that statistical errors and biases towards the initial conditions will be identical in the two cases, thus making the comparison of the resulting ensembles more robust. In general, this can be of critical importance when intensive MD simulations are involved. In fact, in case of a variation in the reference experimental data, such as the availability of additional data, the procedure can be easily renewed using the same prior ensemble, with no necessity of repeating the MD simulations \cite{orioli2020learn}.
The usage of approximate back-calculated SAXS intensities in the enhanced sampling stage, however, helped us not to incur in the difficult situation where the initial ensemble is so poor that it cannot match experimental data when reweighted \cite{cesari2018using,rangan2018determination}.
Our comparison of the back-calculated SAXS intensities obtained from the simulation of crystal structures
using different ion concentrations suggests that, for a fixed RNA structure,
the effect of the ions present in the solution on the spectrum is limited.
This provides a rationale for our choice to reweight our KCl simulation using both KCl and MgCl data and, indirectly,
justifies the neglection of the 10mM MOPSO present in the original experiment.
We also notice that the reweighted ensembles are relatively poor and show a very limited Kish size. Longer simulations might allow for better converged ensembles. However, the reported calculation is at the limit of what is currently feasible in terms of system size and statistical sampling. In particular, the investigated system is significantly larger (57 nucleotides) than other RNA systems where
maximum-entropy or related methods have been used so far to integrate solution experiments with atomistic simulations \cite{borkar2016structure,borkar2017simultaneous,kooshapur2018structural,paissoni2019martini,bottaro2020integrating,reisser2020conformational} (29 nucleotides at most) and contains more heterogeneous structural motifs.

To the best of our knowledge, this is the first report on a procedure to reconstruct the conformational ensemble of a functional RNA molecule that combines experimental SAXS spectra and MD simulations while fully accounting for the solvent contribution, which is included explicitly in the computation, and RNA conformational dynamics. Previous works have mostly addressed separately the two issues. For instance, a common practice used when computing SAXS spectra with explicit solvent methods is to apply soft position restraints to the backbone atoms of the solute biomolecule \cite{kofinger2013atomic,knight2015waxsis}, although other studies were conducted where unrestrained solutes were examined \cite{hermann2019saxs,chen2015structural}.
If used to analyze relatively short plain MD simulations where, by necessity, the solute dynamics is limited, these methods would mostly report the solvent fluctuations.
An opposite scenario is offered by the pure solute restraints implemented in PLUMED \cite{paissoni2019martini}, that
can be naturally combined with efficient enhanced sampling methods so as to properly characterize solute dynamics
\cite{bonomi2016metadynamic}. In this case, however, solvent contributions are neglected by construction.
In this respect, our protocol takes the best from both worlds to reconstruct conformational ensembles that are compatible with reference experimental SAXS data, by (a) enforcing approximate solute-only spectra during the simulation and (b) reweighting the resulting ensemble using the more accurate explicit solvent models.

\section{ACKNOWLEDGEMENTS}

Sandro Bottaro and Carlo Camilloni are acknowledged for reading the manuscript
and providing useful suggestions.
\bibliographystyle{ieeetr}
\bibliography{main}

\begin{thebibliography}{10}

\bibitem{al2008rna}
H.~M. Al-Hashimi and N.~G. Walter, ``Rna dynamics: it is about time,'' {\em
  Curr. Opin. Struct. Biol.}, vol.~18, no.~3, pp.~321--329, 2008.

\bibitem{mustoe2014hierarchy}
A.~M. Mustoe, C.~L. Brooks, and H.~M. Al-Hashimi, ``Hierarchy of rna functional
  dynamics,'' {\em Annu. Rev. Biochem.}, vol.~83, pp.~441--466, 2014.

\bibitem{dror2012biomolecular}
R.~O. Dror, R.~M. Dirks, J.~Grossman, H.~Xu, and D.~E. Shaw, ``Biomolecular
  simulation: a computational microscope for molecular biology,'' {\em Annu.
  Rev. Biophys.}, vol.~41, pp.~429--452, 2012.

\bibitem{sponer2018rna}
J.~{\v S}poner, G.~Bussi, M.~Krepl, P.~Ban{\'a}{\v s}, S.~Bottaro, R.~A. Cunha,
  A.~Gil-Ley, G.~Pinamonti, S.~Poblete, P.~Jure\v{c}ka, N.~G. Walter, and
  M.~Otyepka, ``{RNA} structural dynamics as captured by molecular simulations:
  a comprehensive overview,'' {\em Chem. Rev.}, vol.~118, no.~8,
  pp.~4177--4338, 2018.

\bibitem{pitera2012use}
J.~W. Pitera and J.~D. Chodera, ``On the use of experimental observations to
  bias simulated ensembles,'' {\em J. Chem. Theory Comput.}, vol.~8, no.~10,
  pp.~3445--3451, 2012.

\bibitem{bonomi2016metainference}
M.~Bonomi, C.~Camilloni, A.~Cavalli, and M.~Vendruscolo, ``Metainference: A
  bayesian inference method for heterogeneous systems,'' {\em Sci. Adv.},
  vol.~2, no.~1, p.~e1501177, 2016.

\bibitem{bonomi2017principles}
M.~Bonomi, G.~T. Heller, C.~Camilloni, and M.~Vendruscolo, ``Principles of
  protein structural ensemble determination,'' {\em Curr. Opin. Struct. Biol.},
  vol.~42, pp.~106--116, 2017.

\bibitem{bottaro2018biophysical}
S.~Bottaro and K.~Lindorff-Larsen, ``Biophysical experiments and biomolecular
  simulations: A perfect match?,'' {\em Science}, vol.~361, no.~6400,
  pp.~355--360, 2018.

\bibitem{cesari2018using}
A.~Cesari, S.~Rei{\ss}er, and G.~Bussi, ``Using the maximum entropy principle
  to combine simulations and solution experiments,'' {\em Computation}, vol.~6,
  no.~1, p.~15, 2018.

\bibitem{borkar2016structure}
A.~N. Borkar, M.~F. Bardaro, C.~Camilloni, F.~A. Aprile, G.~Varani, and
  M.~Vendruscolo, ``Structure of a low-population binding intermediate in
  protein-rna recognition,'' {\em Proc. Natl. Acad. Sci. USA}, vol.~113,
  no.~26, pp.~7171--7176, 2016.

\bibitem{borkar2017simultaneous}
A.~N. Borkar, P.~Vallurupalli, C.~Camilloni, L.~E. Kay, and M.~Vendruscolo,
  ``Simultaneous nmr characterisation of multiple minima in the free energy
  landscape of an rna uucg tetraloop,'' {\em Physical chemistry chemical
  physics}, vol.~19, no.~4, pp.~2797--2804, 2017.

\bibitem{kooshapur2018structural}
H.~Kooshapur, N.~R. Choudhury, B.~Simon, M.~M{\"u}hlbauer, A.~Jussupow,
  N.~Fernandez, A.~N. Jones, A.~Dallmann, F.~Gabel, C.~Camilloni, {\em et~al.},
  ``Structural basis for terminal loop recognition and stimulation of
  pri-mirna-18a processing by hnrnp a1,'' {\em Nat. Commun.}, vol.~9, no.~1,
  pp.~1--17, 2018.

\bibitem{paissoni2019martini}
C.~Paissoni, A.~Jussupow, and C.~Camilloni, ``Martini bead form factors for
  nucleic acids and their application in the refinement of protein--nucleic
  acid complexes against saxs data,'' {\em J. Appl. Crystallogr.}, vol.~52,
  no.~2, pp.~394--402, 2019.

\bibitem{bottaro2020integrating}
S.~Bottaro, P.~J. Nichols, B.~V{\"o}geli, M.~Parrinello, and
  K.~Lindorff-Larsen, ``Integrating nmr and simulations reveals motions in the
  uucg tetraloop,'' {\em Nucleic Acids Res.}, vol.~48, no.~11, pp.~5839--5848,
  2020.

\bibitem{reisser2020conformational}
S.~Rei{\ss}er, S.~Zucchelli, S.~Gustincich, and G.~Bussi, ``Conformational
  ensembles of an rna hairpin using molecular dynamics and sparse nmr data,''
  {\em Nucleic Acids Res.}, vol.~48, no.~3, pp.~1164--1174, 2020.

\bibitem{chen2016saxs}
Y.~Chen and L.~Pollack, ``Saxs studies of rna: structures, dynamics, and
  interactions with partners,'' {\em Wiley Interdiscip. Rev. RNA}, vol.~7,
  no.~4, pp.~512--526, 2016.

\bibitem{hub2018interpreting}
J.~S. Hub, ``Interpreting solution x-ray scattering data using molecular
  simulations,'' {\em Curr. Opin. Struct. Biol.}, vol.~49, pp.~18--26, 2018.

\bibitem{kofinger2013atomic}
J.~K{\"o}finger and G.~Hummer, ``Atomic-resolution structural information from
  scattering experiments on macromolecules in solution,'' {\em Phys. Rev. E},
  vol.~87, no.~5, p.~052712, 2013.

\bibitem{chen2015structural}
P.-c. Chen and J.~S. Hub, ``Structural properties of protein--detergent
  complexes from saxs and md simulations,'' {\em J. Phys. Chem. Lett.}, vol.~6,
  no.~24, pp.~5116--5121, 2015.

\bibitem{hermann2019saxs}
M.~R. Hermann and J.~S. Hub, ``Saxs-restrained ensemble simulations of
  intrinsically disordered proteins with commitment to the principle of maximum
  entropy,'' {\em J. Chem. Theory Comput.}, vol.~15, no.~9, pp.~5103--5115,
  2019.

\bibitem{paissoni2020determination}
C.~Paissoni, A.~Jussupow, and C.~Camilloni, ``Determination of protein
  structural ensembles by hybrid-resolution saxs restrained molecular
  dynamics,'' {\em J. Chem. Theory Comput.}, vol.~16, no.~4, pp.~2825--2834,
  2020.

\bibitem{jussupow2020dynamics}
A.~Jussupow, A.~C. Messias, R.~Stehle, A.~Geerlof, S.~M. Solbak, C.~Paissoni,
  A.~Bach, M.~Sattler, and C.~Camilloni, ``The dynamics of linear
  polyubiquitin,'' {\em Sci. Adv.}, vol.~6, no.~42, p.~eabc3786, 2020.

\bibitem{ivanovic2020small}
M.~T. Ivanović, M.~R. Hermann, M.~Wójcik, J.~Pérez, and J.~S. Hub,
  ``Small-angle x-ray scattering curves of detergent micelles: Effects of
  asymmetry, shape fluctuations, disorder, and atomic details,'' {\em J. Phys.
  Chem. Lett.}, vol.~11, no.~3, pp.~945--951, 2020.

\bibitem{svergun1995crysol}
D.~Svergun, C.~Barberato, and M.~H. Koch, ``Crysol--a program to evaluate x-ray
  solution scattering of biological macromolecules from atomic coordinates,''
  {\em J. Appl. Crystallogr.}, vol.~28, no.~6, pp.~768--773, 1995.

\bibitem{schneidman2010foxs}
D.~Schneidman-Duhovny, M.~Hammel, and A.~Sali, ``Foxs: a web server for rapid
  computation and fitting of saxs profiles,'' {\em Nucleic Acids Res.},
  vol.~38, no.~suppl\_2, pp.~W540--W544, 2010.

\bibitem{nguyen2014accurate}
H.~T. Nguyen, S.~A. Pabit, S.~P. Meisburger, L.~Pollack, and D.~A. Case,
  ``Accurate small and wide angle x-ray scattering profiles from atomic models
  of proteins and nucleic acids,'' {\em J. Chem. Phys.}, vol.~141, no.~22,
  p.~12B608\_1, 2014.

\bibitem{grishaev2010improved}
A.~Grishaev, L.~Guo, T.~Irving, and A.~Bax, ``Improved fitting of solution
  x-ray scattering data to macromolecular structures and structural ensembles
  by explicit water modeling,'' {\em J. Am. Chem. Soc.}, vol.~132, no.~44,
  pp.~15484--15486, 2010.

\bibitem{knight2015waxsis}
C.~J. Knight and J.~S. Hub, ``Waxsis: a web server for the calculation of
  saxs/waxs curves based on explicit-solvent molecular dynamics,'' {\em Nucleic
  Acids Res.}, vol.~43, no.~W1, pp.~W225--W230, 2015.

\bibitem{marchi2016first}
M.~Marchi, ``A first principle particle mesh method for solution saxs of large
  bio-molecular systems,'' {\em J. Chem. Phys.}, vol.~145, no.~4, p.~045101,
  2016.

\bibitem{park2009simulated}
S.~Park, J.~P. Bardhan, B.~Roux, and L.~Makowski, ``Simulated x-ray scattering
  of protein solutions using explicit-solvent models,'' {\em J. Chem. Phys.},
  vol.~130, no.~13, p.~04B607, 2009.

\bibitem{chen2009molecular}
A.~A. Chen, D.~E. Draper, and R.~V. Pappu, ``Molecular simulation studies of
  monovalent counterion-mediated interactions in a model rna kissing loop,''
  {\em J. Mol. Biol.}, vol.~390, no.~4, pp.~805--819, 2009.

\bibitem{ivanovic2018quantifying}
M.~T. Ivanovi{\'c}, L.~K. Bruetzel, R.~Shevchuk, J.~Lipfert, and J.~S. Hub,
  ``Quantifying the influence of the ion cloud on saxs profiles of charged
  proteins,'' {\em Phys. Chem. Chem. Phys.}, vol.~20, no.~41, pp.~26351--26361,
  2018.

\bibitem{mlynsky2018exploring}
V.~Ml{\`y}nsk{\`y} and G.~Bussi, ``Exploring rna structure and dynamics through
  enhanced sampling simulations,'' {\em Curr. Opin. Struct. Biol.}, vol.~49,
  pp.~63--71, 2018.

\bibitem{moazed1988interaction}
D.~Moazed, J.~M. Robertson, and H.~F. Noller, ``Interaction of elongation
  factors ef-g and ef-tu with a conserved loop in 23s rna,'' {\em Nature},
  vol.~334, no.~6180, pp.~362--364, 1988.

\bibitem{gao2009structure}
Y.-G. Gao, M.~Selmer, C.~M. Dunham, A.~Weixlbaumer, A.~C. Kelley, and
  V.~Ramakrishnan, ``The structure of the ribosome with elongation factor g
  trapped in the posttranslocational state,'' {\em Science}, vol.~326,
  no.~5953, pp.~694--699, 2009.

\bibitem{weixlbaumer2008insights}
A.~Weixlbaumer, H.~Jin, C.~Neubauer, R.~M. Voorhees, S.~Petry, A.~C. Kelley,
  and V.~Ramakrishnan, ``Insights into translational termination from the
  structure of rf2 bound to the ribosome,'' {\em Science}, vol.~322, no.~5903,
  pp.~953--956, 2008.

\bibitem{welty2018divalent}
R.~Welty, S.~A. Pabit, A.~M. Katz, G.~D. Calvey, L.~Pollack, and K.~B. Hall,
  ``Divalent ions tune the kinetics of a bacterial gtpase center rrna folding
  transition from secondary to tertiary structure,'' {\em RNA}, vol.~24,
  no.~12, pp.~1828--1838, 2018.

\bibitem{conn2002compact}
G.~L. Conn, A.~G. Gittis, E.~E. Lattman, V.~K. Misra, and D.~E. Draper, ``A
  compact rna tertiary structure contains a buried backbone--k+ complex,'' {\em
  J. Mol. Biol.}, vol.~318, no.~4, pp.~963--973, 2002.

\bibitem{welty2020ribosomal}
R.~Welty, M.~Rau, S.~Pabit, M.~S. Dunstan, G.~L. Conn, L.~Pollack, and K.~B.
  Hall, ``Ribosomal protein l11 selectively stabilizes a tertiary structure of
  the gtpase center rrna domain,'' {\em J. Mol. Biol.}, vol.~432, no.~4,
  pp.~991--1007, 2020.

\bibitem{cornell1995second}
W.~D. Cornell, P.~Cieplak, C.~I. Bayly, I.~R. Gould, K.~M. Merz, D.~M.
  Ferguson, D.~C. Spellmeyer, T.~Fox, J.~W. Caldwell, and P.~A. Kollman, ``A
  second generation force field for the simulation of proteins, nucleic acids,
  and organic molecules,'' {\em J. Am. Chem. Soc.}, vol.~117, no.~19,
  pp.~5179--5197, 1995.

\bibitem{perez2007refinement}
A.~P{\'e}rez, I.~March{\'a}n, D.~Svozil, J.~Sponer, T.~E. Cheatham~III, C.~A.
  Laughton, and M.~Orozco, ``Refinement of the amber force field for nucleic
  acids: improving the description of $\alpha$/$\gamma$ conformers,'' {\em
  Biophys. J.}, vol.~92, no.~11, pp.~3817--3829, 2007.

\bibitem{zgarbova2011refinement}
M.~Zgarbov{\'a}, M.~Otyepka, J.~Šponer, A.~Mládek, P.~Banáš, T.~E.
  Cheatham~III, and P.~Jurecka, ``Refinement of the cornell et al. nucleic
  acids force field based on reference quantum chemical calculations of
  glycosidic torsion profiles,'' {\em J. Chem. Theory Comput.}, vol.~7, no.~9,
  pp.~2886--2902, 2011.

\bibitem{izadi2014building}
S.~Izadi, R.~Anandakrishnan, and A.~V. Onufriev, ``Building water models: a
  different approach,'' {\em J. Phys. Chem. Lett.}, vol.~5, no.~21,
  pp.~3863--3871, 2014.

\bibitem{Joung2008determination}
I.~S. Joung and T.~E. Cheatham~III, ``Determination of alkali and halide
  monovalent ion parameters for use in explicitly solvated biomolecular
  simulations,'' {\em J. Phys. Chem. B}, vol.~112, no.~30, pp.~9020--9041,
  2008.

\bibitem{allner2012magnesium}
O.~Alln{\'e}r, L.~Nilsson, and A.~Villa, ``Magnesium ion--water coordination
  and exchange in biomolecular simulations,'' {\em J. Chem. Theory Comput.},
  vol.~8, no.~4, pp.~1493--1502, 2012.

\bibitem{best2014balanced}
R.~B. Best, W.~Zheng, and J.~Mittal, ``Balanced protein--water interactions
  improve properties of disordered proteins and non-specific protein
  association,'' {\em J. Chem. Theory Comput.}, vol.~10, no.~11,
  pp.~5113--5124, 2014.

\bibitem{palazzesi2015accuracy}
F.~Palazzesi, M.~K. Prakash, M.~Bonomi, and A.~Barducci, ``Accuracy of current
  all-atom force-fields in modeling protein disordered states,'' {\em J. Chem.
  Theory Comput.}, vol.~11, no.~1, pp.~2--7, 2015.

\bibitem{shabane2019general}
P.~S. Shabane, S.~Izadi, and A.~V. Onufriev, ``General purpose water model can
  improve atomistic simulations of intrinsically disordered proteins,'' {\em J.
  Chem. Theory Comput.}, vol.~15, no.~4, pp.~2620--2634, 2019.

\bibitem{bergonzo2015improved}
C.~Bergonzo and T.~E. Cheatham~III, ``Improved force field parameters lead to a
  better description of rna structure,'' {\em J. Chem. Theory Comput.},
  vol.~11, no.~9, pp.~3969--3972, 2015.

\bibitem{bottaro2018conformational}
S.~Bottaro, G.~Bussi, S.~D. Kennedy, D.~H. Turner, and K.~Lindorff-Larsen,
  ``Conformational ensembles of rna oligonucleotides from integrating nmr and
  molecular simulations,'' {\em Sci. Adv.}, vol.~4, no.~5, p.~eaar8521, 2018.

\bibitem{abraham2015gromacs}
M.~J. Abraham, T.~Murtola, R.~Schulz, S.~P{\'a}ll, J.~C. Smith, B.~Hess, and
  E.~Lindahl, ``Gromacs: High performance molecular simulations through
  multi-level parallelism from laptops to supercomputers,'' {\em SoftwareX},
  vol.~1, pp.~19--25, 2015.

\bibitem{bussi2007canonical}
G.~Bussi, D.~Donadio, and M.~Parrinello, ``Canonical sampling through velocity
  rescaling,'' {\em J. Chem. Phys.}, vol.~126, no.~1, p.~014101, 2007.

\bibitem{parrinello1981polymorphic}
M.~Parrinello and A.~Rahman, ``Polymorphic transitions in single crystals: A
  new molecular dynamics method,'' {\em J. Appl. Phys.}, vol.~52, no.~12,
  pp.~7182--7190, 1981.

\bibitem{sugita2000replica}
Y.~Sugita and Y.~Okamoto, ``Replica-exchange multicanonical algorithm and
  multicanonical replica-exchange method for simulating systems with rough
  energy landscape,'' {\em Chem. Phys. Lett.}, vol.~329, no.~3-4, pp.~261--270,
  2000.

\bibitem{bussi2014hamiltonian}
G.~Bussi, ``Hamiltonian replica exchange in gromacs: a flexible
  implementation,'' {\em Mol. Phys.}, vol.~112, no.~3-4, pp.~379--384, 2014.

\bibitem{laio2002escaping}
A.~Laio and M.~Parrinello, ``Escaping free-energy minima,'' {\em Proc. Natl.
  Acad. Sci. USA}, vol.~99, no.~20, pp.~12562--12566, 2002.

\bibitem{camilloni2008exploring}
C.~Camilloni, D.~Provasi, G.~Tiana, and R.~A. Broglia, ``Exploring the protein
  g helix free-energy surface by solute tempering metadynamics,'' {\em
  Proteins}, vol.~71, no.~4, pp.~1647--1654, 2008.

\bibitem{bussi2020using}
G.~Bussi and A.~Laio, ``Using metadynamics to explore complex free-energy
  landscapes,'' {\em Nat. Rev. Phys.}, vol.~2, no.~4, pp.~200--212, 2020.

\bibitem{bottaro2014role}
S.~Bottaro, F.~Di~Palma, and G.~Bussi, ``The role of nucleobase interactions in
  rna structure and dynamics,'' {\em Nucleic Acids Res.}, vol.~42, no.~21,
  pp.~13306--13314, 2014.

\bibitem{bottaro2016free}
S.~Bottaro, P.~Ban{\'a}{\v s}, J.~{\v S}poner, and G.~Bussi, ``Free energy
  landscape of gaga and uucg rna tetraloops,'' {\em J. Phys. Chem. Lett.},
  vol.~7, no.~20, pp.~4032--4038, 2016.

\bibitem{barducci2008well}
A.~Barducci, G.~Bussi, and M.~Parrinello, ``Well-tempered metadynamics: a
  smoothly converging and tunable free-energy method,'' {\em Phys. Rev. Lett.},
  vol.~100, no.~2, p.~020603, 2008.

\bibitem{ferrarotti2015accurate}
M.~J. Ferrarotti, S.~Bottaro, A.~P{\'e}rez-Villa, and G.~Bussi, ``Accurate
  multiple time step in biased molecular simulations,'' {\em J. Chem. Theory
  Comput.}, vol.~11, no.~1, pp.~139--146, 2015.

\bibitem{tribello2014plumed}
G.~A. Tribello, M.~Bonomi, D.~Branduardi, C.~Camilloni, and G.~Bussi, ``Plumed
  2: New feathers for an old bird,'' {\em Comput. Phys. Commun.}, vol.~185,
  no.~2, pp.~604--613, 2014.

\bibitem{bonomi2019promoting}
M.~Bonomi, G.~Bussi, C.~Camilloni, G.~A. Tribello, P.~Ban{\'a}{\v{s}},
  A.~Barducci, M.~Bernetti, P.~G. Bolhuis, S.~Bottaro, D.~Branduardi, {\em
  et~al.}, ``Promoting transparency and reproducibility in enhanced molecular
  simulations,'' {\em Nat. Methods}, vol.~16, no.~8, pp.~670--673, 2019.

\bibitem{branduardi2012metadynamics}
D.~Branduardi, G.~Bussi, and M.~Parrinello, ``Metadynamics with adaptive
  gaussians,'' {\em J. Chem. Theory Comput.}, vol.~8, no.~7, pp.~2247--2254,
  2012.

\bibitem{gonzalez2019quality}
R.~González-Alemán, D.~Hernández-Castillo, J.~Caballero, and L.~A.
  Montero-Cabrera, ``Quality threshold clustering of molecular dynamics: a word
  of caution,'' {\em J. Chem. Inf. Mod.}, vol.~60, no.~2, pp.~467--472, 2019.

\bibitem{virtanen2020scipy}
P.~Virtanen, R.~Gommers, T.~E. Oliphant, M.~Haberland, T.~Reddy, D.~Cournapeau,
  E.~Burovski, P.~Peterson, W.~Weckesser, J.~Bright, {\em et~al.}, ``Scipy 1.0:
  fundamental algorithms for scientific computing in python,'' {\em Nat.
  Methods}, vol.~17, no.~3, pp.~261--272, 2020.

\bibitem{gray1969survey}
P.~G. Gray and L.~Kish, ``Survey sampling.,'' {\em J. Royal Stat. Soc. A
  (General)}, vol.~132, no.~2, p.~272, 1969.

\bibitem{rangan2018determination}
R.~Rangan, M.~Bonomi, G.~T. Heller, A.~Cesari, G.~Bussi, and M.~Vendruscolo,
  ``Determination of structural ensembles of proteins: restraining vs
  reweighting,'' {\em J. Chem. Theory Comput.}, vol.~14, no.~12,
  pp.~6632--6641, 2018.

\bibitem{schneidman2013accurate}
D.~Schneidman-Duhovny, M.~Hammel, J.~A. Tainer, and A.~Sali, ``Accurate saxs
  profile computation and its assessment by contrast variation experiments,''
  {\em Biophys. J.}, vol.~105, no.~4, pp.~962--974, 2013.

\bibitem{kimanius2015saxs}
D.~Kimanius, I.~Pettersson, G.~Schluckebier, E.~Lindahl, and M.~Andersson,
  ``Saxs-guided metadynamics,'' {\em J. Chem. Theory Comput.}, vol.~11, no.~7,
  pp.~3491--3498, 2015.

\bibitem{orioli2020learn}
S.~Orioli, A.~H. Larsen, S.~Bottaro, and K.~Lindorff-Larsen, ``How to learn
  from inconsistencies: Integrating molecular simulations with experimental
  data,'' {\em Prog. Mol. Biol. Transl. Sci.}, vol.~170, pp.~123--176, 2020.

\bibitem{bonomi2016metadynamic}
M.~Bonomi, C.~Camilloni, and M.~Vendruscolo, ``Metadynamic metainference:
  Enhanced sampling of the metainference ensemble using metadynamics,'' {\em
  Sci. Rep.}, vol.~6, no.~1, pp.~1--11, 2016.

\end{thebibliography}

\end{document}